\documentclass[conference]{IEEEtran}
\IEEEoverridecommandlockouts
\usepackage{cite}
\usepackage{amsmath,amssymb,amsfonts}
\usepackage{algorithmic}
\usepackage{numprint}
\usepackage{graphicx}
\usepackage{textcomp}
\usepackage{multirow}
\usepackage{makecell}
\usepackage{setspace}
\usepackage{array}
\usepackage{subcaption}
\usepackage{booktabs} 
\usepackage{siunitx} 
\usepackage{caption}
\usepackage{adjustbox}
\usepackage[table]{xcolor}
\usepackage{xspace}
\usepackage{url}

\def\BibTeX{{\rm B\kern-.05em{\sc i\kern-.025em b}\kern-.08em
    T\kern-.1667em\lower.7ex\hbox{E}\kern-.125emX}}

\usepackage[most]{tcolorbox}

\tcbset{
    orangebox/.style={
        breakable,
        colback=orange!20,
        colframe=orange!50,
        width=\columnwidth,
        boxrule=0.5pt,
        arc=4pt,
        left=4pt, right=4pt, top=2pt, bottom=2pt
    }
}

\tcbset{
    lightgraybox/.style={
        breakable,
        colback=olive!15,
        colframe=olive!50,
        width=\columnwidth,
        boxrule=0.5pt,
        arc=4pt,
        left=4pt, right=4pt, top=2pt, bottom=2pt
    }
}

\definecolor{greenArrow}{RGB}{0, 128, 0}
\definecolor{redArrow}{RGB}{255, 0, 0}

\newcommand{\upArrow}{\textcolor{greenArrow}{↑}}
\newcommand{\downArrow}{\textcolor{redArrow}{↓}}
\newcommand{\tool}{\textsc{VulTegra}\xspace}

\begin{document}

\title{
It Only Gets Worse: Revisiting DL-Based Vulnerability Detectors from a Practical Perspective
}

\author{
    \IEEEauthorblockN{
        Yunqian Wang$^\ast$, Xiaohong Li$^\ast$, 
        Yao Zhang, Yuekang Li, Zhiping Zhou, Ruitao Feng$^\dagger$
    }
    \IEEEauthorblockA{
        \textit{$^1$College of Intelligence and Computing, Tianjin University, China} \\
        \textit{$^2$University of New South Wales, Australia} \\
        \textit{$^3$The Southern Cross University, Australia}
    }
    \thanks{$^\ast$These authors contributed equally to this work.}
    \thanks{$^\dagger$Corresponding author.}
}

\maketitle

\begin{abstract}
With the escalating threat of software vulnerabilities to the security of modern software systems, an increasing number of deep learning (DL) model-based vulnerability detectors have been developed for vulnerability detection. Despite these advancements, the consistency of these detectors within their declared Common Weakness Enumeration (CWE) ranges, their real-world effectiveness, and the applicability of different model types across various scenarios all remain uncertain. This uncertainty may lead to unreliable detection results in practical applications, increased false positives and false negatives, and limited adaptability to newly emerged vulnerabilities. Conducting a large-scale and in-depth analysis of DL-based vulnerability detectors can help uncover critical factors influencing detection performance, improve the design and training of these models, and enhance their practical deployment in real-world scenarios. 

In this paper, we present \tool, a novel evaluation framework that, for the first time, conducts a multidimensional assessment comparing scratch-trained models and pre-trained-based models for vulnerability detection, while verifying key factors influencing detection performance. Our framework reveals that state-of-the-art (SOTA) detectors still suffer from low consistency, limited practical detection capabilities, and limited scalability. Moreover, comparative results indicate that the increasingly favored pre-trained-based models are not universally superior to scratch-trained models; instead, they exhibit distinct strengths and application scenarios. Most importantly, our study highlights the limitations of relying solely on CWE-based classification and reveals a set of critical factors that significantly influence detection performance. Experimental validation shows that these factors have a substantial impact: modifying only any single factor led to recall improvements across all seven evaluated SOTA detectors, with six detectors also achieving higher F1 scores. Our findings provide deep insights into model behavior, highlighting the need to consider both vulnerability types and inherent code features.
\end{abstract}

\section{Introduction}
In recent years, the number of disclosed vulnerabilities recorded in the Common Vulnerabilities and Exposures (CVE) database\cite{cve_website} has steadily increased, with a growing proportion of these vulnerabilities classified as high-risk. This trend underscores the urgent need for effective and reliable vulnerability detection tools to mitigate security risks and protect software systems from exploitation\cite{ref1}.

Traditional vulnerability detection relies on static analysis\cite{staticanalysis1, staticanalysis2}, dynamic analysis\cite{dynamicanalysis1,dynamicanalysis2,dynamicanalysis3}, and symbolic execution\cite{symbolicexecution}. While effective in certain cases, these techniques often struggle with scalability, high false positive rates, and unseen vulnerability patterns. DL-based vulnerability detection has emerged as a promising alternative\cite{ref2}, broadly divided into scratch-trained models and pre-trained-based models. Scratch-trained models are built from the ground up using vulnerability-specific datasets, enabling them to capture tailored patterns but requiring high-quality labeled data for effective training\cite{ref3}. In contrast, pre-trained-based models are initialized with general knowledge learned from large-scale code corpora, allowing them to leverage broad representations that can aid in detecting unseen vulnerabilities\cite{scale}. In addition, it is worth mentioning that although large language models (LLMs) have developed rapidly in recent years and performed well in various fields, existing studies\cite{ref4} suggest that LLMs still fall short of specialized SOTA vulnerability detectors in the domain of vulnerability detection. Therefore, we only evaluate DL-based vulnerability detectors in this study.

Despite the success of both scratch-trained models and pre-trained-based models for vulnerability detection, our understanding of these detectors remains incomplete. Chakraborty et al.\cite{ref5} evaluated the capability of DL-based vulnerability detectors on realistic datasets, but their study overlooked whether these detectors could consistently maintain high accuracy within their declared CWE\cite{cwe} ranges or effectively detect newly emerging vulnerabilities. Moreover, their work did not explore the differences between scratch-trained models and pre-trained-based models. Steenhoek et al.\cite{ref6} conducted evaluation work that focused on both types of detectors, but they did not perform a comprehensive comparative analysis from a multidimensional perspective. Additionally, most current vulnerability detection research relies on CWE as the sole criterion for sample selection\cite{ref7}. However, the CWE classification standard does not account for the intrinsic code characteristics of vulnerability samples. To the best of our knowledge, no research has explored whether there are hidden factors beyond the CWE classification that affect vulnerability detection performance. Addressing these concerns is crucial for guiding the development and deployment of more effective and adaptable vulnerability detection models in real-world practice.

Therefore, to better understand how different detection model types behave in practical detection tasks, and what factors influence their detection performance, we propose a novel framework \tool. It, for the first time, performs a multidimensional assessment to compare scratch-trained models and pre-trained-based models and verifies key factors affecting detection performance. This framework enables a comprehensive comparison of scratch-trained models and pre-trained-based models for vulnerability detection, providing insights into their consistency, real-world effectiveness, and scalability capabilities, while verifying key factors influencing detection performance.

Leveraging this framework, we conducted an extensive evaluation involving four SOTA scratch-trained models and three SOTA pre-trained-based models for vulnerability detection. Our experimental results reveal several critical insights: (a) Existing detectors are still far from achieving stable and reliable results across their declared vulnerability scope, while the consistency among the seven detectors remains below 80\%. (b) The actual detection capabilities of these detectors in the real world are limited, and further improvements are needed before these detectors can be used for real-world software vulnerability detection tasks. (c) While current detectors perform well on previously seen vulnerabilities, their scalability to unknown vulnerabilities remains a critical gap. (d) Scratch-trained models and pre-trained-based models exhibit distinct strengths and application scenarios. (e) Modifying even a single key factor improved recall across all seven evaluated detectors, with six detectors also showing higher F1 scores. The largest improvements were a 35\% increase in recall and a 13.8\% increase in F1 score.

Our findings offer guidance for future research on DL-based vulnerability detection. By uncovering key factors that influence detection performance and evaluating the strengths and limitations of both scratch-trained models and pre-trained-based models, our work provides a foundation for designing more effective and adaptable detection models. Furthermore, the insights gained from our study can inform the development of improved training data strategies, enhance model generalization to unseen vulnerabilities, and facilitate the deployment of more reliable detectors in practical scenarios.

In summary, our contributions to the community are as follows:

\begin{itemize}
\item We propose a novel framework, \tool\cite{VulTegra}, which enables a multi-dimensional evaluation of SOTA DL-based vulnerability detectors. Utilizing this framework, we identify issues impacting the practical usage of the detectors, such as low detection consistency, limited real-world applicability, and poor scalability.
\item We present the first comparative study between scratch-trained models and pre-trained-based models for vulnerability detection, revealing distinct strengths and weaknesses suitable for varied scenarios. This finding highlights the importance of understanding model-specific characteristics and selecting the appropriate detector based on the deployment requirements of the target vulnerabilities.
\item We introduce a new set of hidden factors overlooked by the current sample selection standard. Our experimental validation demonstrates that these factors significantly influence detection performance, offering a practical pathway for enhancing the effectiveness of DL-based vulnerability detectors.
\end{itemize}
\section{Preliminary}

\subsection{DL-based Vulnerability Detectors}

DL-based vulnerability detectors can be broadly divided into two categories based on their model construction methodology: scratch-trained models and pre-trained-based models.

\textbf{Scratch-trained models} refer to neural networks constructed and trained entirely from scratch using task-specific labeled datasets tailored for vulnerability detection\cite{funded,vuldeelocator,ivdetect}. These models are designed to learn vulnerability-specific patterns directly from the provided data, making them highly adaptable to the characteristics of the training dataset\cite{ref2}. However, their performance is heavily reliant on the quality, quantity, and diversity of the labeled data, and they may exhibit limited generalization capabilities in detecting unseen vulnerabilities\cite{ref9}.

In contrast, \textbf{pre-trained-based models} are built upon neural architectures that have been pre-trained on large-scale, general-purpose source code corpora, such as CodeBERT\cite{codebert}, GraphCodeBERT\cite{graphcodebert}, or UniXcoder\cite{unixcoder}, and subsequently fine-tuned for vulnerability detection tasks. These models leverage the rich contextual and semantic representations acquired during pre-training, which enables them to perform effectively even in scenarios with limited labeled data.

\begin{figure}[htbp]
  \centering
  \includegraphics[width=0.5\textwidth]{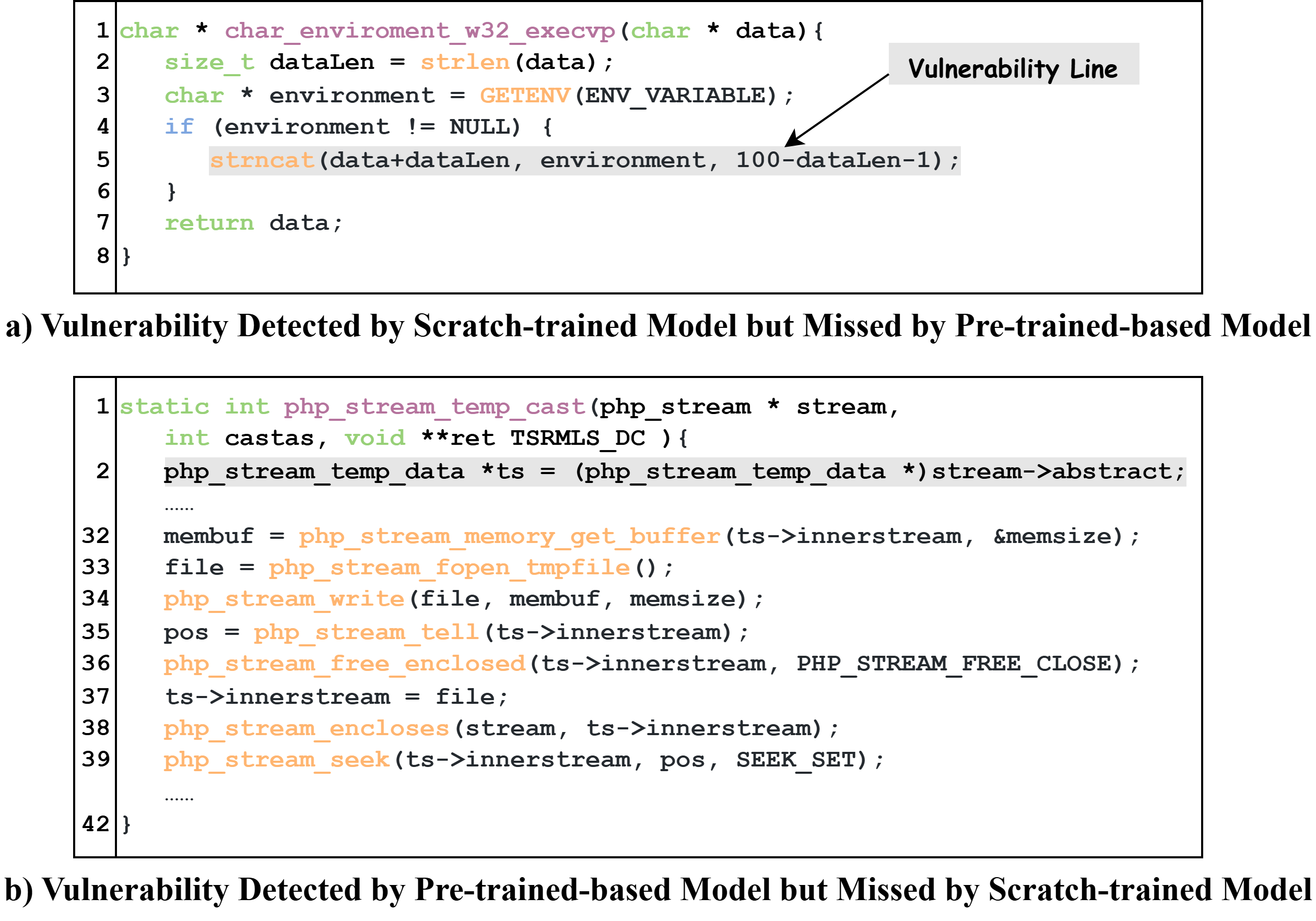}
  \caption{Contrasting Performance of Scratch-Trained Model and Pre-trained-based Model}
  \label{fig:contrasting_performance}
\end{figure}

Despite the rapidly increasing attention toward pre-trained-based models for vulnerability detection, it remains unclear how their performance compares to scratch-trained models across various scenarios. The detection performance of these models can vary significantly depending on code structures and vulnerability contexts. Figure~\ref{fig:contrasting_performance} shows two cases: (a) where a pre-trained-based model successfully identifies a vulnerability that a scratch-trained model fails to detect, and (b) where the scratch-trained model correctly detects a vulnerability missed by the pre-trained-based model. This observation suggests that the performance of detectors is highly context-dependent, and no single model type consistently outperforms the other across all scenarios. Therefore, a comprehensive, multidimensional evaluation of both model types is essential for a deeper understanding of their detection capabilities and practical applicability.

To ensure a fair and consistent evaluation, we carefully selected the vulnerability detectors in our study. We prioritized models that are publicly accessible and well-supported by the research community, ensuring transparency and reproducibility of results. Our focus was on detectors specifically designed or fine-tuned for vulnerability detection tasks that are capable of processing large-scale programming language corpora. Models that are closed-source or not adaptable for fine-tuning, as well as those designed primarily for natural language processing tasks (e.g., LLMs), were excluded to maintain consistency and relevance to the domain of software vulnerability detection.

\begin{tcolorbox}[lightgraybox]
    \textbf{Motivation-1:}
    Existing studies have not fully explored the comparative strengths and limitations of scratch-trained models and pre-trained-based models for vulnerability detection. A comprehensive, multidimensional evaluation is essential to understand their real-world detection capabilities and guide the development of more effective and adaptable vulnerability detection models.
\end{tcolorbox}

\subsection{Current Sample Selection Standard}

Currently, most vulnerability research works select samples based on Common Weakness Enumeration (CWE). CWE provides a hierarchical taxonomy that categorizes vulnerabilities based on their root causes and behaviour patterns, such as buffer overflows (CWE-120), integer overflows (CWE-190), and use-after-free errors (CWE-416). This classification facilitates systematic analysis, dataset annotation, and performance benchmarking for vulnerability detection models\cite{ref10}.

\begin{figure}[htbp]
  \centering
  \includegraphics[width=0.5\textwidth]{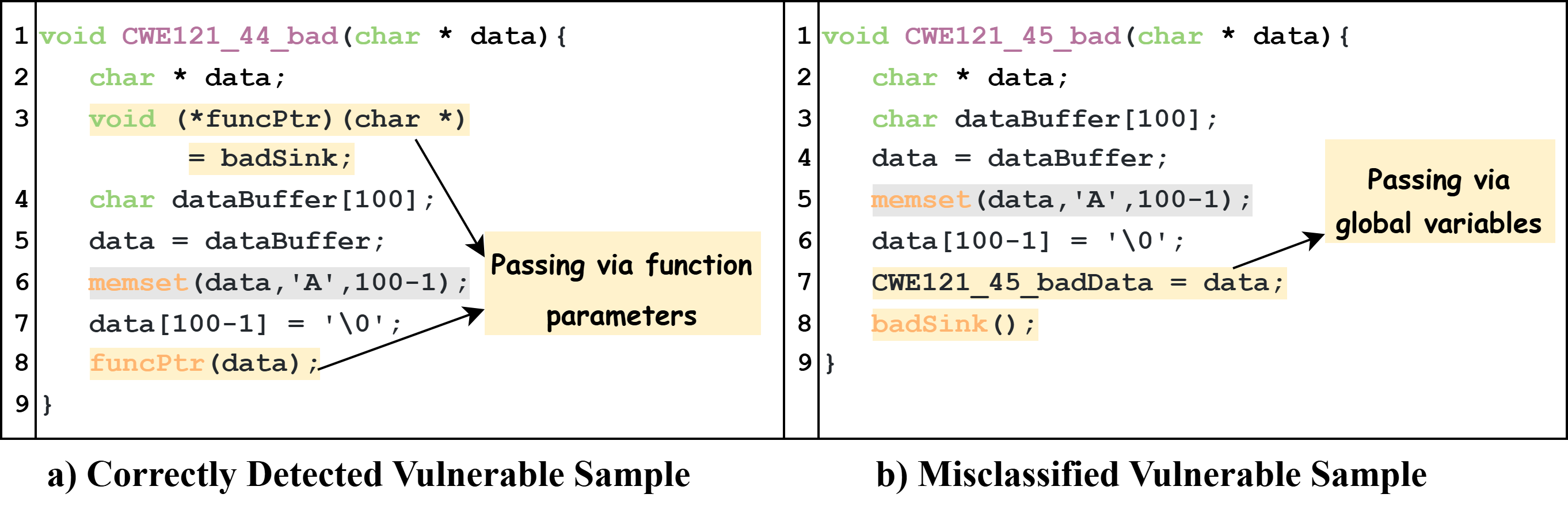}
  \caption{Impact of Data Passing Mechanisms on Detecting Vulnerabilities in CWE-121}
  \label{fig:data_passing}
\end{figure}

However, our early experiments revealed a significant problem of using CWE as the sole basis for sample selection. Even when two samples belong to the same CWE category and share nearly identical structures, they can yield drastically different detection results. For example, as shown in Figure~\ref{fig:data_passing}, two code samples labelled as CWE-121 (Stack-based Buffer Overflow) differ only in the way data is passed. While sample (a) is accurately detected as a vulnerability, sample (b) is misclassified as a non-vulnerable sample. This discrepancy suggests that CWE-based classification may overlook subtle code-level variations influencing detection performance. Finer-grained hidden factors, beyond CWE labels, may play a critical role in generating detection outcomes.

\begin{tcolorbox}[lightgraybox]
    \textbf{Motivation-2:}  
    Relying solely on CWE for sample selection may overlook subtle code-level differences that impact detection performance. Even samples within the same CWE category can yield different results, highlighting the need to explore hidden factors beyond CWE labels to improve vulnerability detection accuracy.
\end{tcolorbox}




\section{Study on DL-based Vulnerability Detectors}

\subsection{Overview}
\begin{figure}[htbp]
  \centering
  \includegraphics[width=0.45\textwidth]{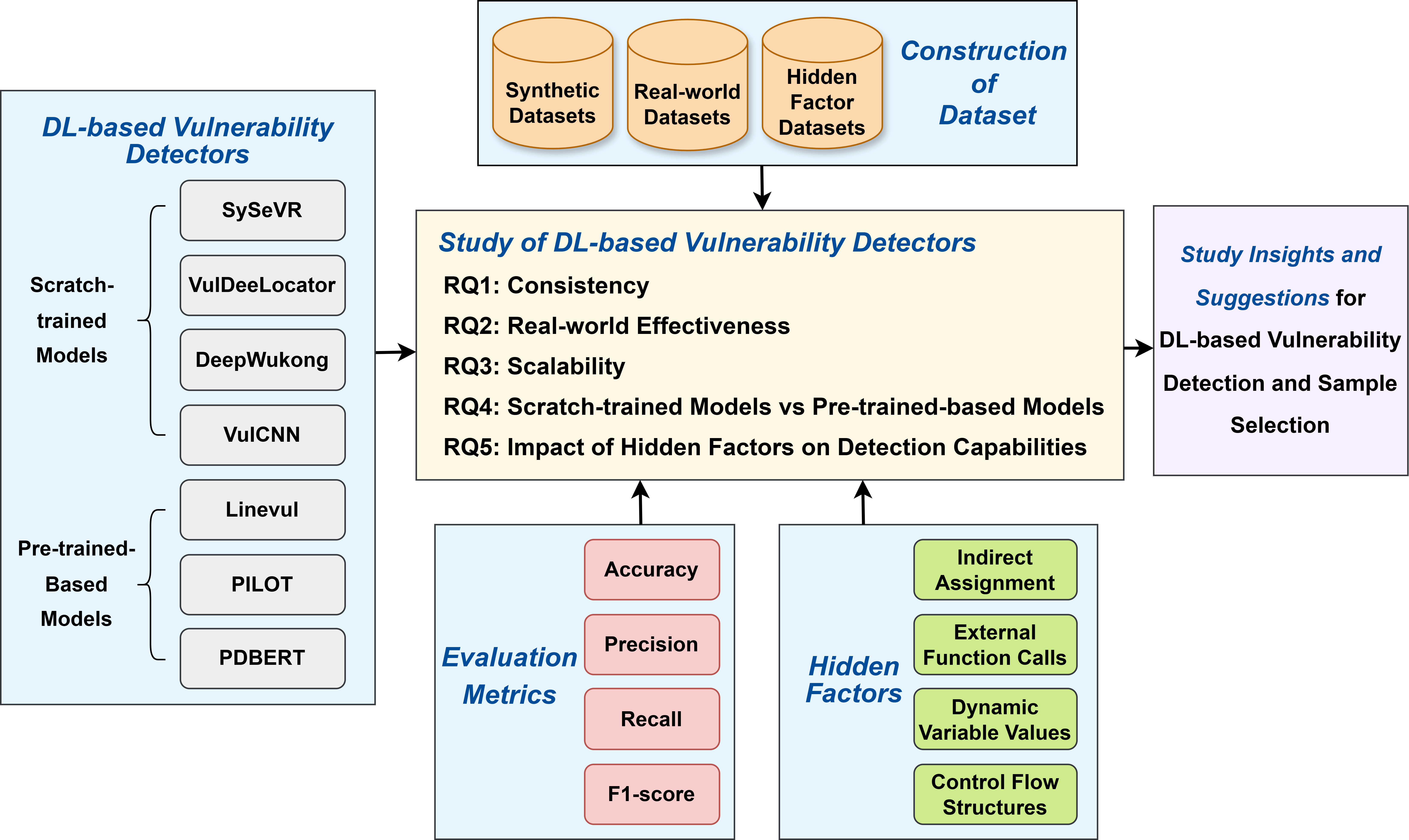}
  \caption{Overview of the Study Framework}
  \label{fig:overview}
\end{figure}

Figure~\ref{fig:overview} provides an overview of the framework employed in our study to comprehensively evaluate and compare DL-based vulnerability detectors, with a particular focus on scratch-trained models and pre-trained-based models. The overarching goal of this study is to produce a deep and systematic understanding of the detection capabilities of state-of-the-art detectors, compare their strengths and weaknesses across multiple dimensions, and explore how hidden factors influence detection performance.

\begin{figure*}[htbp]
  \centering
  \includegraphics[width=\textwidth]{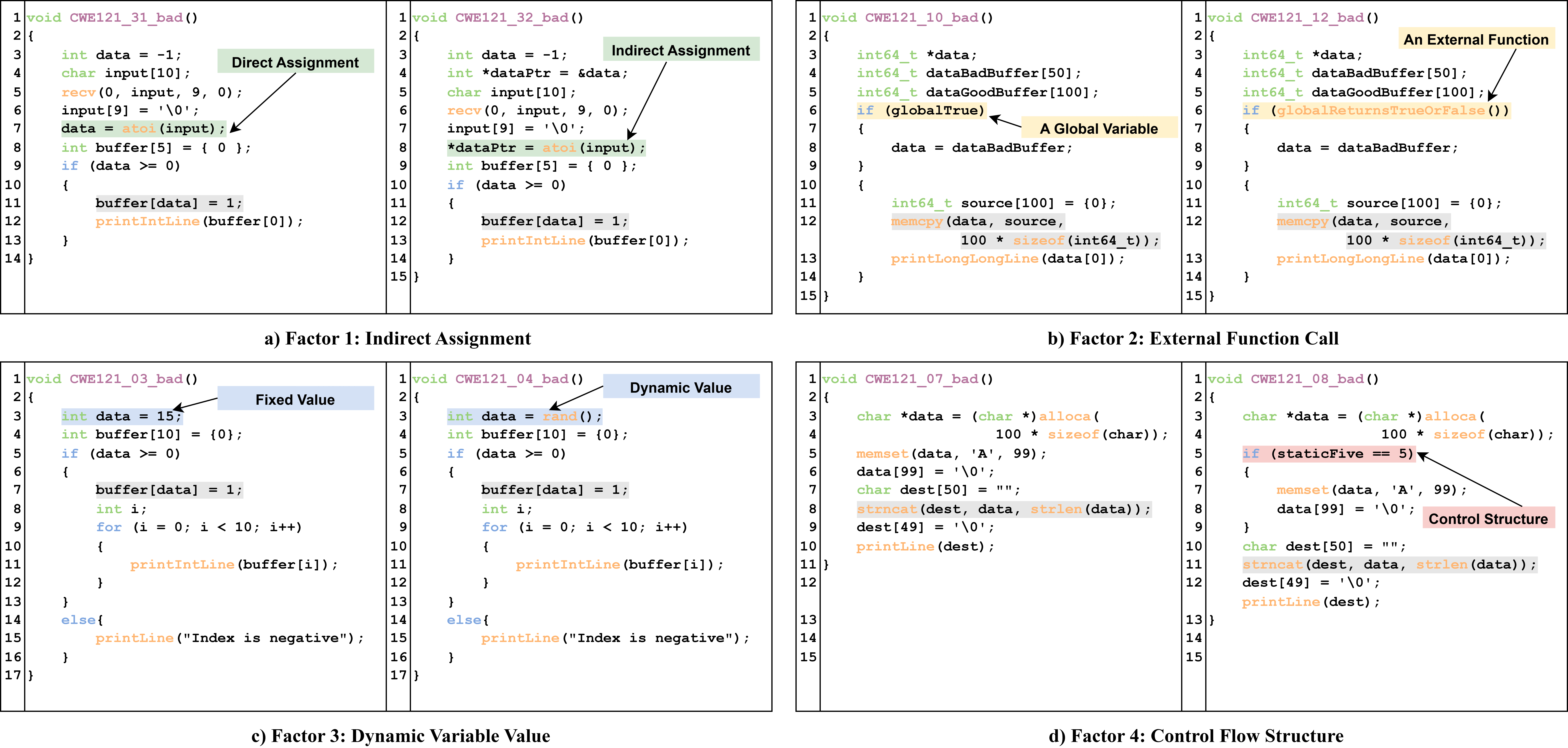}
  \caption{Hidden Factors Affecting Detection Performance}
  \label{fig:hidden_factors}
\end{figure*}


The following five research questions guide our study:
\begin{itemize}
\item \textbf{RQ1: How consistent are DL-based vulnerability detectors across different CWE categories?} This question evaluates whether current detectors can deliver stable and reliable performance within their declared CWE categories.
\item \textbf{RQ2: How effective are DL-based vulnerability detectors in detecting real-world software vulnerabilities?} This problem aims to evaluate the practical effectiveness of DL-based detectors by assessing their performance on real-world vulnerability datasets.
\item \textbf{RQ3: Can DL-based vulnerability detectors effectively detect newly emerged vulnerabilities?} This question examines the scalability and generalization of current detectors when facing previously unseen or newly emerged vulnerability instances.
\item \textbf{RQ4: How do pre-trained-based models vs. scratch-trained models compare in vulnerability detection?} This question aims to compare the detection capabilities of the two mainstream model paradigms and investigate whether the increasingly popular pre-trained-based models consistently outperform their scratch-trained counterparts.
\item \textbf{RQ5: Do hidden factors significantly affect the performance of DL-based vulnerability detectors?} This question verifies whether hidden factors other than CWE labels have a significant impact on the detectors' detection results.
\end{itemize}

By systematically answering these questions, our work aims to fill the current knowledge gaps in the understanding and practical application of DL-based vulnerability detection models. In particular, we conduct a first thorough comparative analysis of scratch-trained models and pre-trained-based models, revealing their respective advantages and suitable application scenarios. Furthermore, we identify and empirically verify a set of key influencing factors that substantially impact detection performance, offering new insights into how detection capabilities can be effectively enhanced.

\subsection{Evaluation Metrics}
The performance of the selected vulnerability detectors is evaluated using the following standard metrics:
\begin{itemize}
\item \textbf{Accuracy:} Measures the overall correctness of the detection model by calculating the proportion of correctly predicted vulnerability samples out of all the samples.
\item \textbf{Precision:} Focuses on the correctness of positive predictions, indicating the proportion of true positive vulnerabilities out of all instances predicted as vulnerable.
\item \textbf{Recall:} Measures the model's ability to detect actual vulnerabilities, indicating the proportion of true positive vulnerabilities out of all actual vulnerabilities.
\item \textbf{F1 Score:} Provides a balanced evaluation between precision and recall, offering a single score to assess the model's overall detection performance.
\end{itemize}

These metrics are essential for comparing the effectiveness of different vulnerability detection models and understanding their reliability across various detection tasks.

\subsection{Hidden Factors}

In our evaluation of vulnerability detectors, we observed inconsistent detection results for samples classified under the same CWE category. Despite exhibiting highly similar code structures, some samples were correctly identified, while others were misclassified. This situation piqued our curiosity, prompting us to manually analyze over 500 samples, carefully comparing the differences between correctly and incorrectly classified cases. Through this detailed analysis, we identified four potential hidden factors that could significantly affect the detectors’ performance: indirect assignment, external function calls, dynamic variable values, and control flow structures. Figure~\ref{fig:hidden_factors} illustrates these factors through a set of comparative code samples.

\textbf{Indirect assignment} refers to cases where data is assigned to a variable through an intermediary operation rather than a direct assignment. Figure~\ref{fig:hidden_factors}(a) shows two nearly identical code snippets, differing only in the way data is passed. Indirect assignment can introduce subtle variations in data flow and memory access patterns, which some detectors may fail to capture accurately.

\textbf{External function call} involves the reliance on functions defined outside the immediate code context. Figure~\ref{fig:hidden_factors}(b) compares two similar samples, where one includes a call to an external function while the other does not. External function calls can obscure the code’s execution path and introduce uncertainty in data flow analysis, leading to inconsistent detection results.

\textbf{Dynamic variable value} pertains to the use of dynamically assigned values at runtime. Figure~\ref{fig:hidden_factors}(c) presents two structurally similar code samples, differing only in the use of dynamic versus static values. Dynamic values introduce variability and increase the complexity of control flow analysis, making it harder for detectors to accurately predict vulnerability states.

\textbf{Control flow structure} reflects the complexity of conditional and loop-based execution paths within a code sample. Figure~\ref{fig:hidden_factors}(d) shows two code snippets that differ in the complexity of their control flow. Complex control structures can obscure vulnerability patterns and confuse detectors, leading to misclassifications.

\subsection{Selected Detectors}
The selected detectors include both scratch-trained models and pre-trained-based models. These models were chosen based on their superior performance (SOTA) in vulnerability detection and their established relevance in the field. The selected detectors, proposed by varied research teams, encompass a wide range of architectures and methodologies, ensuring a thorough evaluation of their strengths and limitations. Table~\ref{tab:vuln_detection} provides an overview of the key characteristics of these detectors.

\begin{table}[ht]
\scriptsize
\centering
\caption{Studied DL-based Vulnerability Detectors}
\label{tab:vuln_detection}
\renewcommand{\arraystretch}{1.2} 
\setlength{\tabcolsep}{3pt} 
\resizebox{\columnwidth}{!}{ 
\begin{tabular}{|c|c|c|c|}
\hline
\textbf{Type} & \textbf{Detector} & \textbf{Architecture} & \textbf{Detection Level} \\ \hline
\multirow{4}{*}{\shortstack{Scratch-trained}} 
& SySeVR & RNN & Function \\ \cline{2-4} 
& VulDeeLocator & Bi-LSTM & Function \\ \cline{2-4} 
& DeepWuKong & GCN & Function \\ \cline{2-4} 
& VulCNN & CNN & Function \\ \hline
\multirow{3}{*}{\shortstack{Pre-trained-based}} 
& LineVul & Transformer & Function/Line \\ \cline{2-4} 
& PILOT & Transformer & Function \\ \cline{2-4} 
& PDBERT & Transformer & Function \\ \hline
\end{tabular}
}
\end{table}

\begin{itemize}
\item \textbf{SySeVR\cite{sysevr}} extracts syntactic and semantic code slices, encoding them to capture vulnerability-related patterns. It was one of the first models to combine both structural and contextual code features, improving detection accuracy.
\item \textbf{VulDeeLocator\cite{vuldeelocator}} applies BiLSTM with attention to locating vulnerable code segments. Its attention mechanism helps focus on critical vulnerability patterns, enhancing detection precision.  
\item \textbf{DeepWuKong\cite{deepwukong}} uses GCN to represent code as graphs, capturing complex data and control flow dependencies. It was among the first models to leverage graph-based learning for vulnerability detection.  
\item \textbf{VulCNN\cite{vulcnn}} applies CNN to tokenized code, automatically extracting hierarchical patterns. Its strength lies in effectively handling structured code vulnerabilities.
\item \textbf{LineVul\cite{linevul}} is a transformer-based model fine-tuned for line-level vulnerability detection. It leverages contextual embeddings to capture subtle vulnerability patterns.
\item \textbf{PILOT\cite{pilot}} builds on CodeBERT and uses masked language modelling to capture both token-level and structural patterns, improving generalization across different vulnerability types.
\item \textbf{PDBERT\cite{pdbert}}, based on BERT, combines syntactic and semantic embeddings to enhance detection accuracy, particularly in complex code scenarios.
\end{itemize}

These detectors are considered SOTA due to their innovative architectures, ability to handle diverse vulnerability patterns, and consistent high performance across multiple CWE categories. They have been widely recognized by multiple research studies as leading approaches in the field of software vulnerability detection\cite{ref11,ref12,ref13,ref14,ref15,ref16}. In this study, we reproduced these detectors using their original datasets and code to ensure fidelity to their reported performance and to facilitate a fair and accurate comparison.

\subsection{Dataset}
\label{sec:dataset}
Our dataset consists of three major components: real-world data, synthetic data, and hidden-factor data. The real-world data primarily sources samples from publicly available datasets, including NVD\cite{nvd}, Devign\cite{devign}, Reveal\cite{reveal}, and Fan\cite{fan}. These datasets provide diverse and authentic vulnerability cases. After collection, we performed extensive preprocessing and cleaning, including deduplication, formatting, and removal of irrelevant samples to ensure consistency and minimize noise during model training. The synthetic data is sourced from the SARD dataset\cite{sard}, which contains samples classified according to the CWE taxonomy. Its primary role is to supplement the real-world dataset by providing samples with explicit CWE labels to meet experimental requirements.

The hidden factor data is designed to explore the limitations of CWE-based classification and identify underlying hidden factors affecting vulnerability detection. It consists of five subsets, created by manually modifying 100 real-world vulnerable samples and 100 non-vulnerable samples based on four hidden factors. The first subset consists of the original, unmodified samples, serving as the baseline for comparison. In the second subset, indirect memory allocations were systematically removed without altering the sample's structure, testing the detector’s sensitivity to memory handling. The third subset eliminates all external function calls while preserving core logic, simplifying the sample to evaluate the impact of external dependencies. In the fourth subset, dynamic variable values are replaced with static values to remove variability related to runtime behaviour. The fifth subset simplifies complex control flow structures to evaluate the detector's sensitivity to logical complexity. This hidden factor dataset enables a deeper investigation into the factors influencing detection outcomes beyond CWE-based classification.

In total, our dataset includes 31,673 vulnerability samples and 208,451 non-vulnerability samples, with 80\% of the data sourced from real-world scenarios. This large-scale dataset ensures a balanced and realistic evaluation environment, enhancing the reliability and generalizability of our experimental results. To address specific RQs, we utilized tailored subsets of this dataset. For RQ1, the dataset encompasses 28 CWE categories, including 4,591 vulnerability samples and 6,732 non-vulnerability samples. For RQ2, the dataset comprises 27,418 vulnerability samples and 203,317 non-vulnerability samples. For RQ3, the dataset includes 20 vulnerability samples across 12 CWE categories and four popular open-source projects, each with over 8,000 forks. RQ4 is based on the analysis of the experimental results from RQ1, RQ2, and RQ3, and thus does not have a separate statistical range. For RQ5, the dataset consists of 500 vulnerability samples and 500 non-vulnerability samples, organized into five experimental sub-datasets. These data scales ensure that each RQ experiment is conducted with sufficient and representative samples to yield robust conclusions.
\section{Experiments and Results}

\subsection{RQ1: Consistency across Different CWE Categories}

\newcolumntype{C}[1]{>{\centering\arraybackslash}p{#1}}

\begin{table*}[htbp]
    \centering
    \caption{Performance of DL-based Vulnerability Detectors on Different CWEs. Values greater than or equal to 0.9, indicating satisfactory performance, are highlighted in cyan, while values below 0.9 are highlighted in red. A dash (–) indicates that the corresponding CWE is not within the evaluation scope and thus has no data available.}
    \label{tab:cwe_performance_comparison}
    \renewcommand{\arraystretch}{1.3} 
    \setlength{\tabcolsep}{2pt} 
    \scriptsize 
    \begin{tabular}{@{}C{2cm} *{7}{C{2.1cm}}@{}}
        \toprule
        \textbf{CWE} & \textbf{SySeVR} & \textbf{VulDeeLocator} & \textbf{DeepWuKong} & \textbf{VulCNN} & \textbf{LineVul} & \textbf{PILOT} & \textbf{PDBert} \\
        \midrule
        CWE-17   &  --    &  --    &  --    &  --    & \cellcolor{pink!40}0.862 &  --    &  --    \\
        CWE-20   & \cellcolor{pink!40}0.790 & \cellcolor{pink!40}0.761 & \cellcolor{pink!40}0.678 &  --    &  --    & \cellcolor{cyan!20}0.900 & \cellcolor{cyan!20}0.929 \\
        CWE-22   & \cellcolor{pink!40}0.781 & \cellcolor{pink!40}0.827 & \cellcolor{pink!40}0.746 &  --    & \cellcolor{pink!40}0.715 & \cellcolor{cyan!20}0.906 & \cellcolor{cyan!20}0.911 \\
        CWE-78   & \cellcolor{pink!40}0.375 & \cellcolor{pink!40}0.618 & \cellcolor{pink!40}0.625 & \cellcolor{pink!40}0.850 &  --    & \cellcolor{pink!40}0.732 & \cellcolor{pink!40}0.805 \\
        CWE-79   &  --    &  --    &  --    &  --    &  --    & \cellcolor{pink!40}0.500 & \cellcolor{pink!40}0.500 \\
        CWE-89   & \cellcolor{pink!40}0.562 & \cellcolor{pink!40}0.605 &  --    &  --    &  --    & \cellcolor{cyan!20}0.913 & \cellcolor{cyan!20}0.928 \\
        CWE-119  & \cellcolor{cyan!20}0.928 & \cellcolor{cyan!20}0.918 & \cellcolor{cyan!20}0.916 &  --    &  --    & \cellcolor{pink!40}0.819 & \cellcolor{pink!40}0.836 \\
        CWE-121  & \cellcolor{pink!40}0.872 & \cellcolor{cyan!20}0.909 &  --    & \cellcolor{pink!40}0.823 &  --    &  --    &  --    \\
        CWE-122  &  --    &  --    &  --    & \cellcolor{pink!40}0.601 &  --    &  --    &  --    \\
        CWE-124  &  --    &  --    &  --    & \cellcolor{cyan!20}0.943 &  --    &  --    &  --    \\
        CWE-125  &  --    &  --    & \cellcolor{pink!40}0.827 &  --    &  --    & \cellcolor{pink!40}0.794 & \cellcolor{pink!40}0.752 \\
        CWE-126  &  --    &  --    &  --    & \cellcolor{pink!40}0.717 &  --    &  --    &  --    \\
        CWE-134  &  --    &  --    &  --    & \cellcolor{cyan!20}1.000 &  --    &  --    &  --    \\
        CWE-190  & \cellcolor{cyan!20}0.932 & \cellcolor{cyan!20}0.946 & \cellcolor{cyan!20}0.951 &  --    &  --    &  --    &  --    \\
        CWE-254  &  --    &  --    & \cellcolor{pink!40}0.763 &  --    & \cellcolor{cyan!20}1.000 &  --    &  --    \\
        CWE-269  &  --    &  --    &  --    &  --    & \cellcolor{pink!40}0.841 &  --    &  --    \\
        CWE-284  &  --    &  --    &  --    &  --    & \cellcolor{cyan!20}1.000 &  --    &  --    \\
        CWE-285  &  --    &  --    &  --    &  --    & \cellcolor{cyan!20}1.000 &  --    &  --    \\
        CWE-311  &  --    &  --    &  --    &  --    & \cellcolor{cyan!20}1.000 &  --    &  --    \\
        CWE-358  &  --    &  --    &  --    &  --    & \cellcolor{cyan!20}1.000 &  --    &  --    \\
        CWE-399  &  --    &  --    & \cellcolor{pink!40}0.853 &  --    &  --    &  --    &  --    \\
        CWE-400  & \cellcolor{pink!40}0.813 & \cellcolor{pink!40}0.856 & \cellcolor{pink!40}0.869 &  --    &  --    &  --    &  --    \\
        CWE-415  &  --    &  --    &  --    &  --    & \cellcolor{cyan!20}1.000 &  --    &  --    \\
        CWE-416  & \cellcolor{pink!40}0.604 & \cellcolor{pink!40}0.647 &  --    &  --    &  --    & \cellcolor{pink!40}0.662 & \cellcolor{pink!40}0.623 \\
        CWE-476  & \cellcolor{pink!40}0.485 & \cellcolor{pink!40}0.592 &  --    &  --    &  --    &  --    &  --    \\
        CWE-617  &  --    &  --    &  --    &  --    & \cellcolor{cyan!20}1.000 &  --    &  --    \\
        CWE-787  &  --    &  --    & \cellcolor{cyan!20}0.902 &  --    &  --    & \cellcolor{pink!40}0.863 & \cellcolor{cyan!20}0.907 \\
        CWE-862  &  --    &  --    &  --    &  --    &  --    & \cellcolor{pink!40}0.000 & \cellcolor{pink!40}0.000 \\
        \bottomrule
    \end{tabular}
\end{table*}

To assess the consistency of the vulnerability detectors, we evaluated their accuracy across specific CWE types. A detector was considered consistent for a particular CWE if its accuracy exceeded 90\% for that vulnerability type. Our experimental design focused on evaluating the performance of each model based on the range of vulnerability types they claimed to handle. All detectors were reproduced using their original datasets and code to ensure a fair and accurate comparison. 

For detectors claiming to handle more than 10 vulnerability types, we focused on evaluating their detection accuracy for the top 10 most prevalent CWE types within the range of CWE types they declared. To ensure consistency and relevance, we ranked the CWE types based on their prevalence in real-world software vulnerabilities, following the CWE TOP25\cite{top25} as a reference. For detectors claiming to handle 10 or fewer vulnerability types, we evaluated their performance across all the vulnerability types they explicitly stated they could detect. For models with no specific claims regarding the range of vulnerabilities they could handle, we restricted the evaluation to the top 10 most common CWE types. If there are insufficient samples for the top 10 CWEs, we will move on to the next CWE until we have a sufficient number of samples for meaningful evaluation.


By structuring the experiments in this way, we ensured that the consistency of each detector was assessed under conditions that closely aligned with real-world vulnerabilities. This also allowed us to test how well these detectors performed on both common vulnerabilities and those within their declared detection scope.


\noindent\textbf{Result:} The experimental results are shown in Table~\ref{tab:cwe_performance_comparison}. Most of the detectors struggled to achieve consistent detection performance on the vulnerabilities they claimed to handle. For instance, SySeVR, one of the detectors with a broad range of declared vulnerabilities, exhibited consistent accuracy above 90\% for only 20\% of the declared vulnerability types. In contrast, LineVul showed better performance, achieving consistent detection for 70\% of the declared vulnerability types. However, even this relatively higher performance indicates that existing models are still far from achieving stable and reliable results across their declared vulnerability scope.

The performance also varied when it came to detecting specific vulnerability types. Certain vulnerabilities, such as CWE-78, remained a significant challenge for many detectors. In our tests, all six models failed to achieve more than 90\% accuracy on this vulnerability type, underscoring a common limitation: current vulnerability detection models lack stable detection capabilities for certain types of vulnerabilities. These recurring detection failures highlight a significant gap in the capability of existing models, particularly for vulnerabilities with more complex or nuanced detection patterns. This indicates that further improvements in model robustness, generalization, and the ability to handle diverse vulnerability types are necessary before these systems can be fully relied upon for real-world vulnerability detection.

\begin{tcolorbox}[orangebox]
    \textbf{Observation-1:}  
    DL-based vulnerability detectors exhibit low consistency across declared vulnerability types, with significant performance gaps, especially for complex vulnerabilities, revealing fundamental limitations in current detection models.
\end{tcolorbox}

\subsection{RQ2: Effectiveness on Real-world Vulnerabilities}

The objective of this experiment is to evaluate the real-world effectiveness of deep learning (DL)-based vulnerability detectors. While many DL-based vulnerability detectors have demonstrated high accuracy on synthetic or curated datasets, their performance on real-world code remains uncertain. This experiment aims to assess whether existing DL-based detectors can effectively detect vulnerabilities in practical scenarios.

To achieve this goal, seven SOTA DL-based vulnerability detectors were evaluated on our real-world vulnerability dataset. We measured the detection performance using three key metrics: Precision, Recall, and F1-score. Precision reflects the detectors' ability to minimize false positives, Recall reflects the detectors' ability to capture true vulnerabilities, and F1-score provides a balanced measure of Precision and Recall. 

\noindent\textbf{Result:} Table~\ref{tab:real_world_performance_comparison} shows the experimental results. It indicates that the overall performance of the seven DL-based detectors was suboptimal. Among the evaluated detectors, PDBert achieved the highest precision at 0.59 and the highest f1-score at 0.56, demonstrating a relatively stronger ability to balance between identifying true positives and minimizing false positives. In terms of Recall, LineVul outperformed other detectors, achieving the highest recall at 0.53. This suggests that LineVul was more effective at identifying true vulnerabilities but at the cost of increased false positives, as indicated by its lower precision and f1-score. The gap between precision and recall across different models reflects the trade-off between capturing more vulnerabilities and minimizing incorrect classifications.  

\begin{table}[!t]
\scriptsize
  \centering
  \caption{Evaluation Results for real-world effectiveness of DL-based Vulnerability Detectors. The report column refers to the indicator data given by the detector authors in their papers. A dash (-) indicates that the paper did not present the data. An asterisk (*) indicates that the result was tested on a real-world dataset.}
  \label{tab:real_world_performance_comparison}
  \setlength{\tabcolsep}{3pt}
  \renewcommand{\arraystretch}{1.2}
  \resizebox{0.48\textwidth}{!}{
    \begin{tabular}{lcccccc}
      \toprule
      \textbf{Detector} & \multicolumn{2}{c}{\textbf{Precision}} & \multicolumn{2}{c}{\textbf{Recall}} & \multicolumn{2}{c}{\textbf{F1-score}} \\
      \cmidrule(lr){2-3} \cmidrule(lr){4-5} \cmidrule(lr){6-7}
                        & \textbf{Report} & \textbf{Test} & \textbf{Report} & \textbf{Test} & \textbf{Report} & \textbf{Test} \\
      \midrule
      SySeVR           & 0.90     & 0.32 & -     & 0.47 & 0.92     & 0.38 \\
      VulDeeLocator    & 0.81*     & 0.35 & -     & 0.52 & 0.80*     & 0.42 \\
      DeepWukong       & -     & 0.43 & -     & 0.52 & 0.90*     & 0.47 \\
      VulCNN           & -     & 0.41 & 0.91     & 0.49 & -     & 0.45 \\
      LineVul          & 0.97*     & 0.56 & 0.86*     & \textbf{0.53} & 0.91*     & 0.54 \\
      PILOT            & 0.54*     & 0.57 & 0.55*     & 0.52 & 0.55*     & 0.54 \\
      PDBert           & -     & \textbf{0.59} & -     & 0.53 & 0.59*     & \textbf{0.56} \\
      \bottomrule
    \end{tabular}
  }
\end{table}

Notably, both PDBert and PILOT, which are pre-trained-based models, achieved results that closely align with their originally claimed performance. This can be attributed to the powerful representation capabilities of pre-trained-based models, which allow them to generalize better across diverse real-world codebases. Their ability to leverage contextual semantics and syntactic patterns learned from large-scale code corpora likely contributes to their robustness and adaptability, even in a real-world testing environment that differs from the original training datasets. Furthermore, these models benefit from transfer learning, enabling them to retain high detection effectiveness without requiring extensive fine-tuning on domain-specific data. The stable performance of PDBert and PILOT underscores the advantage of integrating pre-trained contextual embeddings into vulnerability detection tasks, making them more resilient to domain shifts.

\begin{tcolorbox}[orangebox]
    \textbf{Observation-2:}  
    Existing DL-based vulnerability detectors struggle with real-world code complexity, achieving limited precision and recall. It is possible that pre-trained-based models can provide a scalable and transferable basis for real-world vulnerability detection across different codebases, compared to scratch-trained models.
\end{tcolorbox}

\subsection{RQ3: Scalability to Newly Emerged Vulnerabilities}
To assess the scalability of the selected vulnerability detectors in handling newly disclosed vulnerabilities, we collected a set of 20 CVE vulnerabilities that were disclosed in 2024. These vulnerabilities were carefully selected from widely used open-source projects with more than 8,000 forks, ensuring that they represent security threats in actively maintained and highly influential software systems. By focusing on vulnerabilities from such popular projects, we aimed to evaluate the detectors' ability to handle security issues that have real-world impact.

The seven selected vulnerability detectors were tested on this set of 20 new CVE vulnerabilities. Instead of using traditional metrics like precision and recall, we focused on recording the number of vulnerabilities correctly identified by each detector out of the 20 tested. For each detector, we reported the number of successfully detected vulnerabilities to evaluate how well they generalize to previously unseen vulnerabilities.

\begin{figure}[htbp]
  \centering
  \includegraphics[width=0.45\textwidth]{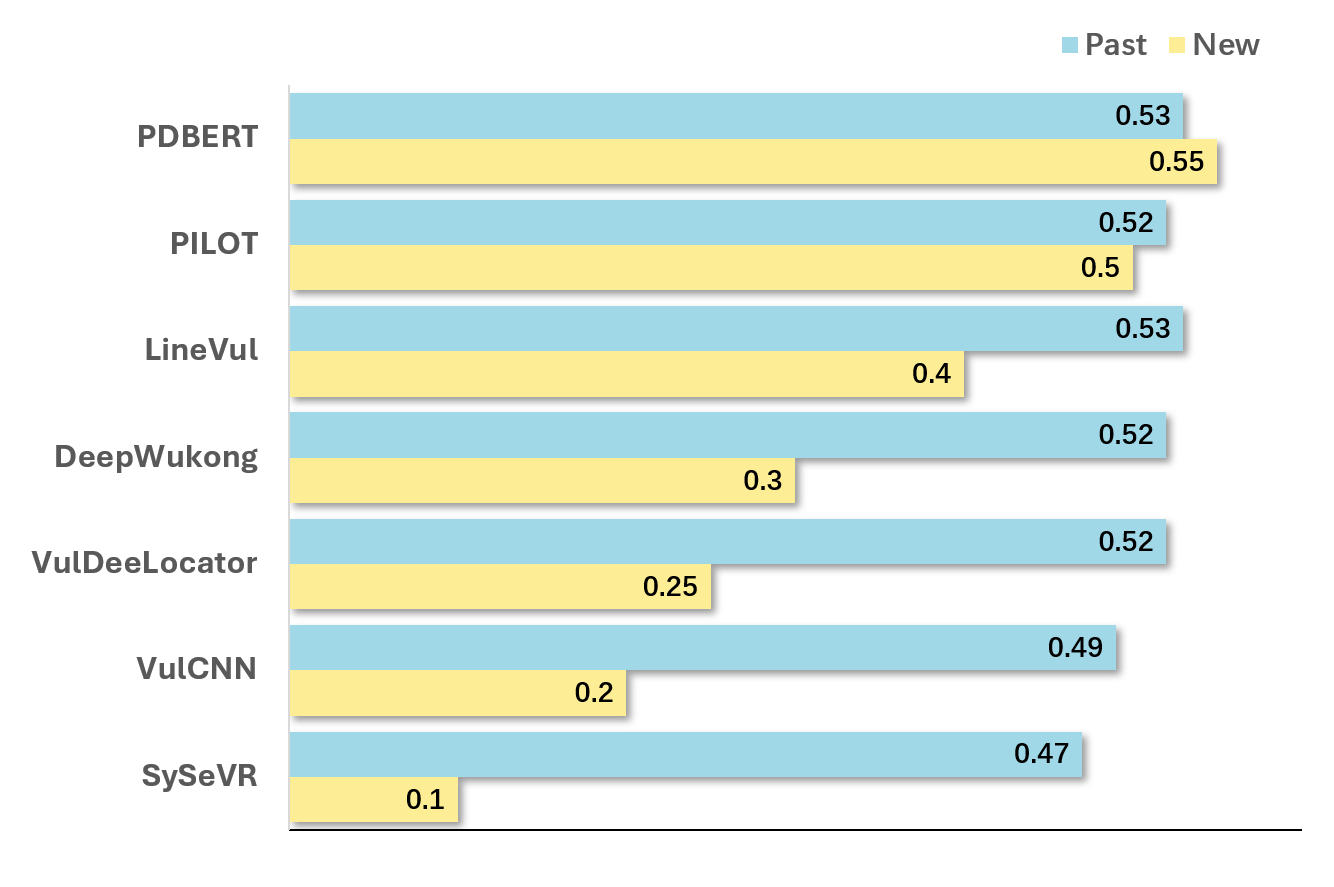}
  \caption{Evaluation Results for Scalability}
  \label{fig:evaluation_scalability}
\end{figure}

\noindent\textbf{Result:} Figure~\ref{fig:evaluation_scalability} shows the experimental results, which revealing that none of the detectors performed well when tested on the newly disclosed 2024 CVE vulnerabilities. The detection accuracy across all models was significantly lower than when tested on the training datasets or on known vulnerabilities.

The SySeVR detector exhibited the lowest detection reliability, with only 10\% accuracy, indicating that it struggles to generalize to new vulnerabilities. This poor performance suggests that SySeVR is limited in its ability to detect emerging vulnerabilities, possibly due to the narrow range of training data it was exposed to.

In contrast, PDBert and PILOT, both pre-trained-based models, achieved relatively higher detection accuracy, with PDBert reaching 55\%. Although this is still insufficient for practical deployment, it is notable that their performance remained close to their results on older datasets, unlike other detectors that suffered severe degradation. This relative stability can be attributed to the rich semantic and syntactic representations learned during pre-training, which enable these models to capture contextual cues beyond surface-level features. Such representations help the models generalize better to previously unseen code structures and vulnerability types, offering greater resilience against data distribution shifts.


\begin{tcolorbox}[orangebox]
    \textbf{Observation-3:}  
    SOTA detectors failed to scale to new vulnerabilities, while PDBert and PILOT retained relatively stable performance, highlighting the superior generalization of pre-trained-based models despite their limitations on unseen threat patterns, compared to scratch-trained models.
\end{tcolorbox}


\subsection{RQ4: Scratch-trained Models vs. Pre-trained-based Models}
The objective of this experiment is to conduct a comprehensive comparison between scratch-trained models and pre-trained-based models for vulnerability detection from the perspectives of consistency, real-world effectiveness, and scalability. While both types of detectors have demonstrated promising performance in vulnerability detection, it remains unclear how their detection capabilities differ across various vulnerability types and practical scenarios. This experiment aims to identify their respective strengths and weaknesses, providing insights into the types of detection tasks each model type is more suited for. To achieve this, we evaluated the performance of four state-of-the-art scratch-trained models and three state-of-the-art pre-trained-based models.

\noindent\textbf{Result:} The experimental results reveal that scratch-trained models and pre-trained-based models have distinct strengths and application scenarios.

According to Table~\ref{tab:cwe_performance_comparison}, we found that pre-trained-based models performed significantly better on vulnerabilities such as CWE-20 (Improper Input Validation) and CWE-89 (SQL Injection), where understanding the semantic context of the code plays a crucial role in detecting these vulnerabilities. On the other hand, scratch-trained models performed better on CWE-119 (Improper Restriction of Operations within the Bounds of a Memory Buffer), a vulnerability that is more closely related to structured code patterns and memory handling. These results suggest that pre-trained-based models excel in scenarios that require contextual understanding of the code, whereas scratch-trained models are more effective in scenarios where vulnerabilities are tied to specific code patterns or structural issues.

As shown in Table~\ref{tab:real_world_performance_comparison}, pre-trained-based models achieved higher precision and F1-scores than scratch-trained models, highlighting their advantage in minimizing false positives and achieving a more balanced detection performance in practical scenarios. Moreover, pre-trained-based models showed stronger scalability, as evidenced by Figure~\ref{fig:evaluation_scalability}, where they demonstrated better adaptability to newly introduced or previously unseen vulnerabilities. This suggests that the broader knowledge base and deeper contextual understanding of pre-trained-based models enable them to generalize more effectively to novel vulnerability patterns.

Overall, these results indicate that scratch-trained models are better suited for detecting vulnerabilities with clear, structured patterns, while pre-trained-based models are more effective for complex and semantic vulnerabilities, and they exhibit greater adaptability to new and evolving vulnerability types.

\begin{tcolorbox}[orangebox]
    \textbf{Observation-4:}  
    Pre-trained-based models excel in detecting complex, semantic vulnerabilities and adapting to new patterns, while scratch-trained models perform better on structured vulnerabilities. This suggests that combining both model types could enhance overall detection performance and adaptability in real-world scenarios.
\end{tcolorbox}

\subsection{RQ5: Impact of Hidden Factors on Detection Capabilities}

To further explore the impact of hidden factors on detection performance, we designed a series of controlled experiments. Our experimental setup involved creating five datasets by manually modifying real-world samples to isolate four hidden factors, namely indirect assignment, external function calls, dynamic variable values, and control flow structures, which were identified through a detailed analysis of over 500 samples (see \S~\ref{sec:dataset} for the details of used data). These modifications were performed by subtracting specific elements from the impacting lines of code based on each factor, thus ensuring controlled variations. We tested seven different vulnerability detectors on this hidden factor dataset, conducting five independent tests for each configuration. Accuracy, precision, recall, and F1-score were used to measure the detectors' performance across different scenarios.

\begin{table}[ht]
\scriptsize
\centering
\caption{The Impact of Key Factors on Detection Performance. \upArrow represents the increase in the metric compared to that in the No-factor dataset, while \downArrow indicates the decrease in the metric relative to that in the No-factor dataset.}
\label{tab:detector_performance_comparison}
\begin{adjustbox}{width=0.48\textwidth}
\begin{tabular}{llcccc}
\toprule
\textbf{Detector} & \textbf{Factor} & \textbf{Acc.} & \textbf{Prec.} & \textbf{Rec.} & \textbf{F1} \\
\midrule

\multirow{5}{*}{\textit{SySeVR}} 
 & No-factor  & 0.300 & 0.308 & 0.320 & 0.314 \\
 & Factor-1   & \upArrow0.335 & \upArrow0.337 & \upArrow0.340 & \upArrow0.338 \\
 & Factor-2   & \upArrow0.360 & \upArrow0.357 & \upArrow0.350 & \upArrow0.353 \\
 & Factor-3   & \upArrow0.365 & \upArrow0.358 & \upArrow0.340 & \upArrow0.349 \\
 & Factor-4   & \upArrow0.340 & \upArrow0.337 & \upArrow0.330 & \upArrow0.333 \\

\midrule
\multirow{5}{*}{\textit{VulDeeLocator}} 
 & No-factor  & 0.330 & 0.327 & 0.340 & 0.333 \\
 & Factor-1   & \upArrow0.355 & \upArrow0.359 & \upArrow0.370 & \upArrow0.364 \\
 & Factor-2   & \upArrow0.390 & \upArrow0.396 & \upArrow0.420 & \upArrow0.408 \\
 & Factor-3   & \upArrow0.385 & \upArrow0.388 & \upArrow0.400 & \upArrow0.394 \\
 & Factor-4   & \upArrow0.370 & \upArrow0.375 & \upArrow0.390 & \upArrow0.382 \\

\midrule
\multirow{5}{*}{\textit{DeepWukong}} 
 & No-factor  & 0.375 & 0.381 & 0.400 & 0.390 \\
 & Factor-1   & \upArrow0.410 & \upArrow0.412 & \upArrow0.420 & \upArrow0.416 \\
 & Factor-2   & \upArrow0.440 & \upArrow0.442 & \upArrow0.460 & \upArrow0.451 \\
 & Factor-3   & \upArrow0.470 & \upArrow0.467 & \upArrow0.430 & \upArrow0.448 \\
 & Factor-4   & \upArrow0.425 & \upArrow0.426 & \upArrow0.420 & \upArrow0.423 \\

\midrule
\multirow{5}{*}{\textit{VulCNN}} 
 & No-factor  & 0.350 & 0.353 & 0.360 & 0.356 \\
 & Factor-1   & \upArrow0.375 & \upArrow0.379 & \upArrow0.390 & \upArrow0.384 \\
 & Factor-2   & \upArrow0.415 & \upArrow0.416 & \upArrow0.420 & \upArrow0.418 \\
 & Factor-3   & \upArrow0.400 & \upArrow0.402 & \upArrow0.410 & \upArrow0.406 \\
 & Factor-4   & \upArrow0.390 & \upArrow0.388 & \upArrow0.380 & \upArrow0.384 \\

\midrule
\multirow{5}{*}{\textit{LineVul}} 
 & No-factor  & 0.505 & 0.667 & 0.020 & 0.575 \\
 & Factor-1   & \upArrow0.530 & \downArrow0.607 & \upArrow0.170 & \downArrow0.566 \\
 & Factor-2   & \upArrow0.640 & \upArrow0.804 & \upArrow0.370 & \upArrow0.713 \\
 & Factor-3   & \upArrow0.585 & \upArrow0.793 & \upArrow0.230 & \upArrow0.673 \\
 & Factor-4   & \downArrow0.400 & \downArrow0.400 & \upArrow0.280 & \downArrow0.400 \\

\midrule
\multirow{5}{*}{\textit{PILOT}} 
 & No-factor  & 0.445 & 0.443 & 0.430 & 0.436 \\
 & Factor-1   & \upArrow0.480 & \upArrow0.480 & \upArrow0.470 & \upArrow0.470 \\
 & Factor-2   & \upArrow0.530 & \upArrow0.532 & \upArrow0.500 & \upArrow0.520 \\
 & Factor-3   & \upArrow0.485 & \upArrow0.485 & \upArrow0.470 & \upArrow0.477 \\
 & Factor-4   & \upArrow0.500 & \upArrow0.500 & \upArrow0.450 & \upArrow0.474 \\

\midrule
\multirow{5}{*}{\textit{PDBert}} 
 & No-factor  & 0.460 & 0.459 & 0.450 & 0.454 \\
 & Factor-1   & \upArrow0.485 & \upArrow0.485 & \upArrow0.470 & \upArrow0.477 \\
 & Factor-2   & \upArrow0.540 & \upArrow0.542 & \upArrow0.520 & \upArrow0.531 \\
 & Factor-3   & \upArrow0.510 & \upArrow0.510 & \upArrow0.490 & \upArrow0.500 \\
 & Factor-4   & \upArrow0.517 & \upArrow0.515 & \upArrow0.460 & \upArrow0.486 \\

\bottomrule
\end{tabular}
\end{adjustbox}
\end{table}

\noindent\textbf{Result:} The experimental results are presented in Table~\ref{tab:detector_performance_comparison}. It is observed that the modification of any single key factor led to an increase in the recall rate for all seven evaluated SOTA detectors. Moreover, six of these detectors also achieved higher F1-scores. Among them, Linevul demonstrated the most significant improvements, with a 35\% increase in recall and a 13.8\% increase in the F1-score. This indicates that the presence of these hidden factors was indeed impeding the detectors' ability to accurately identify vulnerable samples, and their removal or simplification enhanced the detectors' performance in terms of detecting true positives.

In addition, it became evident that external function calls and dynamic variable values had a more substantial impact on vulnerability detection results, while the removal of indirect assignment and complex control flow structures produced mixed results. For instance, LineVul showed a decline in precision and F1-scores when tested on the dataset with indirect assignments removed and the dataset with complex control flow structures simplified. This suggests that while simplifying indirect assignments and control flow structures can improve recall in most cases, it may introduce challenges for some detectors in maintaining a balance between precision and recall, highlighting the complex interplay between these hidden factors and the detectors' underlying algorithms.

Overall, these results underscore the importance of considering these hidden factors when developing and evaluating vulnerability detectors, as they can significantly influence the detectors' performance and accuracy in real - world scenarios.

\begin{tcolorbox}[orangebox]
    \textbf{Observation-5:}  
    Hidden factors such as external function calls, dynamic variable values, indirect assignment, and control flow structures significantly influence detection performance, with the highest increase in recall being 35\% and the highest F1-score increase being 13.8\%. Incorporating these fine-grained code characteristics into classification and detection models could enhance consistency and reliability in real-world scenarios.
\end{tcolorbox}



\section{Discussion \& Threat to Validity}

\subsection{Discussion}

Our study underscores several challenges and opportunities in the field of vulnerability detection models. While DL-based vulnerability detectors show promise, there are key areas where further refinement is necessary to make them more reliable and adaptable for real-world applications.

\subsubsection{Consistency and Scalability of Existing Detectors} One of the most prominent challenges we observed is the inconsistency and limited scalability of existing detectors. While some models demonstrate promising results on specific vulnerability types, they still struggle to maintain consistent performance across the broader range of vulnerabilities encountered in real-world software systems. These models are not yet ready for direct deployment in practical scenarios without further optimization, particularly in terms of generalization across diverse vulnerability types and the ability to handle large-scale, complex software projects.

\subsubsection{Reasonable Use of Two Models to Improve Detection Capabilities} The differences between scratch-trained models and pre-trained-based models suggest that selecting models based on task characteristics is crucial for improving detection performance. Scratch-trained models are better suited for identifying structured, pattern-based vulnerabilities, while pre-trained-based models are more effective in detecting complex, semantic vulnerabilities. Rather than relying on a single model type, combining both can enhance overall accuracy and adaptability.

\subsubsection{Importance of Sample Code Features} Even within a single CWE category, different sample codes may have distinct characteristics, which can lead to varying detection outcomes. This variability in code features can significantly impact the performance of detection models, making it crucial to consider not only the categorization of vulnerabilities but also the specific features of the sample code when evaluating detection models. 

\subsubsection{Future Work} Building on these insights, future research could prioritize two key directions to advance vulnerability detection. First, we can develop hybrid detection techniques that effectively integrate the strengths of scratch-trained models and pre-trained-based models, combining their capabilities to address the diverse demands of real-world software systems. This could involve designing frameworks that employ both model types in tandem, such as using scratch-trained models to detect structured vulnerabilities and pre-trained-based models for semantic analysis, with mechanisms like ensemble voting or layered architectures to optimize overall performance. Second, we can enhance dataset construction by emphasizing hidden factors that reflect code features. By systematically incorporating these factors into training and evaluation datasets, we can improve the training effectiveness of models and ensure a more comprehensive and fair assessment of their detection capabilities.

\subsection{Threat to Validity}
Despite our efforts to ensure a comprehensive and realistic evaluation, several potential threats to validity may influence our findings.

\subsubsection{Data Representativeness} To closely approximate real-world scenarios, we constructed datasets with rich and diverse samples, most of which are sourced from real-world projects and publicly disclosed vulnerabilities. However, these datasets may still fall short of capturing the full diversity and complexity of production environments, including variations in programming styles, domain-specific code patterns, and system architectures.

\subsubsection{Factor Generalizability} To ensure the generality and representativeness of the identified hidden factors, we conducted manual analysis on over 500 vulnerability samples to extract four key factors influencing detection performance. While these factors demonstrated significant impact across multiple detectors and datasets, their applicability may vary in other codebases or detection scenarios. Different programming languages, vulnerability types, or development paradigms may introduce new influencing factors not captured in our current study.

\section{Related Work}
Throughout the course of our research, we have reviewed several relevant studies in the field of vulnerability detection. In this section, we provide a summary of the remaining ones and compare them with our contribution.

\textbf{Vulnerability Detection Techniques:} Vulnerability detection techniques have evolved significantly over time, progressing from traditional static analysis methods to deep learning-based approaches. Static analysis\cite{Cppcheck,SonarQube,DependencyCheck} serves as one of the earliest and most widely used techniques, aiming to identify potential vulnerabilities by analyzing source code without executing it. Although effective for early detection, such techniques often face challenges including limited scalability, high false positive rates, and poor adaptability to unknown vulnerability patterns\cite{ref17,ref18,ref19}. In contrast, deep learning-based techniques offer a data-driven alternative and can be broadly categorized into scratch-trained models and pre-trained-based models. Scratch-trained models learn vulnerability-specific patterns from labeled datasets using architectures like CNNs or hybrid neural networks\cite{train1,train2,train3,train4,train5,train6,train7}. Pre-trained-based models are first trained on large-scale code corpora to capture general code semantics and then fine-tuned for vulnerability detection tasks\cite{pre1,pre2,pre3,pre4}. Compared to traditional methods, deep learning-based techniques provide improved generalization, scalability, and detection accuracy, marking a significant shift in the development of automated vulnerability detection systems.

\textbf{Empirical Analysis of Vulnerability Detection:} Several prior studies have highlighted the limitations of DL-based vulnerability detection models, particularly in real-world applications\cite{ref4,ref6,ref13,ref20}. Despite initial success in controlled environments, these models often struggle with low accuracy and generalization to new, unseen codebases, as they tend to rely on superficial patterns rather than true vulnerability characteristics. Chakraborty et al.\cite{ref5} further demonstrate that the performance of these models significantly drops in practical settings, resulting in high false positive and false negative rates. Similarly, Ding et al.\cite{ref21} show that Code LMs, even with improvements in fine-tuning or prompting, fail to meet the performance demands for realworld vulnerability detection. These challenges underline the need for novel training approaches. Our work builds on these insights by focusing on the multidimensional capabilities of existing models and exploring the impact of subtle variations in code features.

\section{Conclusion}
This work presents an in-depth analysis of the current state of DL-based vulnerability detectors. Our study reveals that current detectors suffer from low consistency, limited real-world effectiveness, and poor scalability to newly emerged vulnerabilities. Our comparative analysis showed that pre-trained-based models, while generally more robust, are not universally superior, and both scratch-trained models and pre-trained-based models exhibit distinct strengths depending on the application scenario. Most importantly, our study identified a set of key factors beyond CWE classification that significantly influence detection performance. These findings offer practical guidance for developing more robust, generalizable, and effective vulnerability detection models.


\bibliographystyle{IEEEtran}
\bibliography{ref}

\end{document}